\documentclass[twocolumn,astrosymb,times]{aastex631}
\usepackage{rotating}
\usepackage{booktabs}
\usepackage{textcomp, gensymb}
\usepackage{tabularx}
\usepackage{amsmath}
\shortauthors{P\'erez Paolino et al.}

\begin{document}

\title{Starspots as an Explanation for the Mysterious IYJ Continuum Excess Emission in Classical T~Tauri Stars}

\correspondingauthor{Facundo P\'erez Paolino}
\email{fperezpa@caltech.edu}

\author[0000-0002-4128-7867]{Facundo P\'erez Paolino}
\affiliation{Colgate University, 13 Oak Drive, Hamilton, NY 13346, USA}
\affiliation{Department of Astronomy, California Institute of Technology, Pasadena, CA 91125, USA}

\author[0000-0001-8642-5867]{Jeffrey S. Bary}
\affiliation{Colgate University, 13 Oak Drive, Hamilton, NY 13346, USA}

\author{Lynne A. Hillenbrand}
\affiliation{Department of Astronomy, California Institute of Technology, Pasadena, CA 91125, USA}

\author[0009-0000-9670-2194]{Madison Markham}
\affiliation{Colgate University, 13 Oak Drive, Hamilton, NY 13346, USA}

\author[0000-0002-3747-2496]{William Fischer}
\affiliation{{Space Telescope Science Institute, 3700 San Martin Drive, Baltimore, MD 21218, USA}}
\affiliation{Visiting Astronomer at the NASA Infrared Telescope Facility}

\altaffiliation{NASA's Infrared Telescope Facility is operated by the University of Hawaii under contract 80HQTR19D0030 with the National Aeronautics and Space Administration.}

\begin{abstract}
An accurate estimation of the continuum excess emission from accretion spots and inner circumstellar disk regions is crucial for a proper derivation of fundamental stellar parameters in accreting systems. However, the presence of starspots can make disentangling the complicated multi-component emission in these systems challenging. Subtraction of a single-temperature spectral template is insufficient to account for the composite stellar emission, as we demonstrated in a recent campaign involving Weak-Lined T Tauri Stars. Here, we model the moderate resolution near-infrared spectra of Classical T Tauri Stars, presenting new spectral models that incorporate spotted stars plus emission from accretion hot-spots and a warm inner disk, allowing us to simultaneously reconstruct the entire {$0.8-2.4~\mu$m} spectrum of our sixteen targets. Using these models, we re-derive the continuum excess emission. Our results indicate that accounting for starspots resolves the need to include a previously proposed intermediate temperature component in the IYJ excess, and highlights the importance of a proper treatment of starspots in studies of accreting low-mass stars.
\end{abstract}


\section{Introduction} \label{sec:intro}
Pre-Main-Sequence (PMS) stars undergo a phase of tumultuous change as they feed from their circumstellar disk through a process of magnetically driven accretion on timescales lasting a few million years \citep[e.g.,][]{Hillenbrand2008}. According to the current magnetospheric accretion paradigm \citep{Hartmann2016}, circumstellar material is channeled toward the forming star through accretion funnels until it crashes onto the stellar surface at nearly free-fall velocities in high-latitude regions distributed around the star. The accreting gas forms shock fronts at the base of the stellar photosphere, heating the infalling gas. This phenomenon is characterized by the presence of broad emission lines (e.g., Ca II, He II, and C IV) in the stellar spectra \citep{Edwards1994,Muzzerolle1998, Ardilla2015}, and by ultraviolet (UV) and $U$-band excesses \citep{Calvet1998}. Due to the high temperature of the accretion shock ($\approx10^6$ K), it emits soft X-rays that are then partially reprocessed by the pre-shock and the stellar photosphere \citep{Kastner2002, kastner2006}. The re-processed radiation then heats up the photosphere below the shock, resulting in optically-thick excess continuum emission from the heated photosphere with typical temperatures between $6000-10000$~K, and with sizes commonly characterized by filling factors under 15\% of the stellar surface area \citep{Long2011}. The accretion funnel and shocks also produce Balmer and Paschen line emission \citep{Muzerolle2003}, for which empirical relationships have been developed to convert line luminosities into mass accretion rates \citep[e.g.,][]{Natta2006}. 

Even if a young star is not actively accreting, it is still characterized by strong magnetic fields that when combined with rapid rotation rates can lead to strong magnetic flares, heightened chromospheric and coronal activity, and the large scale suppression of convection \citep{Berdyugina2005, Toriumi2019}. The formation of cool starspots covering large portions of the stellar surface  \citep{Rydgren&Vrba1983} when combined with the excess emission from accretion shocks and warm circumstellar disks complicates modeling of the Near-Infrared (NIR) spectra of Classical T Tauri Stars (CTTS). However, disentangling the multi-component emission is essential to accurately determine fundamental stellar parameters such as temperature, luminosity, and mass and to develop an accurate understanding of PMS stellar evolution. This process of extracting the excess non-stellar emission from the observed spectrum requires an accurate understanding of the underlying spectrum of the stellar photosphere and the visual extinction \citep{Basri1990, Hartigan1995, Espaillat2010}. 

A common method to accomplish this consists of using a template star of similar spectral type to the target star, and assuming that the photospheric lines in the target star are weaker due to the excess continuum emission filling in, or "veiling," these features \citep[e.g.,][]{Hartigan1991}. For this purpose, spectra of non-accreting Weak-lined T-Tauri Stars are often used as photospheric templates, since they provide better matches to CTTS in surface gravity and chromospheric activity than main sequence dwarfs \citep{Valenti1993, McClure2013,Manara2013}. Once veiling has been measured, the fraction of the observed flux coming from the star itself and the visual extinction $A_{V}$ can be determined. {In order for this procedure to provide meaningful results, two assumptions must be true: 1) the WTTS template is not actively accreting, as that would introduce continuum veiling to the template, and 2) 
the SED of the template star and the target star must be well-matched. Making this second assumption can be particularly challenging, given that many WTTS display evidence for starspots, including periodic modulation of their lightcurves \citep{Grankin1998}, and color excesses \citep{Pecaut2016}. Both of these effects suggest that commonly used template WTTS might be spotted, altering their SEDs, particularly in the NIR, and biasing these measurements.}

\citet{Fischer2011} present a NIR study of 16 actively accreting CTTS in Taurus-Auriga in which they used WTTS templates to derive the excess emission. After subtracting the WTTS templates to remove the spectra of the photospheres, the remaining excess emission was fit with multi-component blackbody spectra, finding the fits to require three separate components in order to reproduce the NIR excesses across all stars in their sample. In addition to excess emission from accretion onto the star (8000~K) and warm dust in the inner disk (1400~K), these results suggest the existence of a third excess component with temperatures ranging between 2200~K and 5000~K, and with filling factors between 0.14 and 11 times the projected stellar surface area. Since this third component did not fit neatly into the model of excess accretion and disk emission, the authors proposed two new plausible origins for this excess: 1) emission from warm annuli encircling the accretion hot spots, or 2) optically-thick gas located inside the dust sublimation radius, accretion flow, or wind. 

In this work, we revisit the \citet{Fischer2011} study by modeling the underlying spectra of the accreting objects as that of a heavily spotted young star. Using BTSettl-CIFIST model atmospheres, we created a grid of spotted star models to fit and then remove the stellar contribution to the NIR spectra of the stars. We show that once the presence of cool starspots is taken into account, the resulting excess spectra can be fit with just two components, accretion and disk, obviating the need for the mysterious third component. {These results highlight the observational efficiency of medium-resolution spectra when combined with simple composite models of stellar surfaces for determining precise stellar continua and veiling in accreting systems.}


\section{The Fischer (2011) SpeX Data}

The spectra analyzed in this paper were collected at NASA's Infrared Telescope Facility (IRTF) using SpeX, a medium-resolution NIR cross-dispersed spectrograph covering 0.8-2.45~$\mu$m at a resolving power of $R\approx2000$. The observations were made on UT 2006 November 26 and 27, which is prior to the SpeX upgrade in January 2014 that increased the spectral coverage. Details of the observations and data reduction are provided in \citet{Fischer2011}, where the data were originally presented.


\section{Spectroscopic Models of Spotted Classical T Tauri Stars}\label{sec:models}
\begin{figure*}
\centering
\includegraphics[width=1\linewidth]{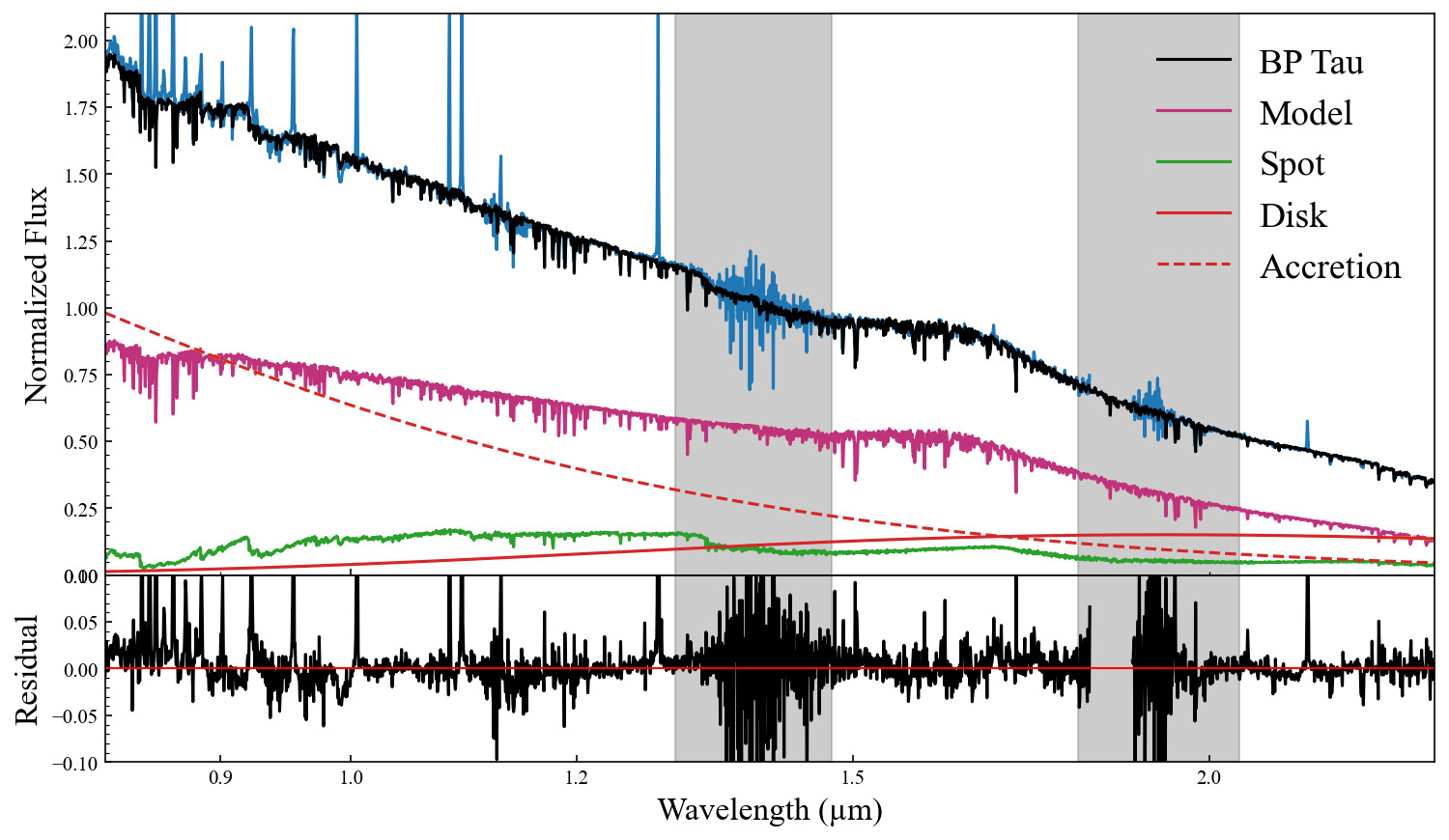}
\caption{{Accreting spotted fit for BP~Tau}. Plotted is a de-reddened BP~Tau spectrum (blue solid line) along with its best-fit spotted accreting model (black solid line). Overplotted are the individual accretion ($f_{acc}=0.068$, $T_{acc}=6736~K$ in dashed red), disk ($f_{disk}=9.24$ in solid red), photospheric ($T_{phot}=3939~K$ in purple), and starspot ($T_{spot}=2699~K$, $f_{spot}=0.51$ in green) components that make up the fit with their proper scaling intact. Below are plotted fractional residuals, with an RMS of 0.029. Regions shaded in grey were masked during the fitting process.} 
\label{fig:bp_tau_fit}
\end{figure*}

\begin{figure*}
\centering
\includegraphics[width=1\linewidth]{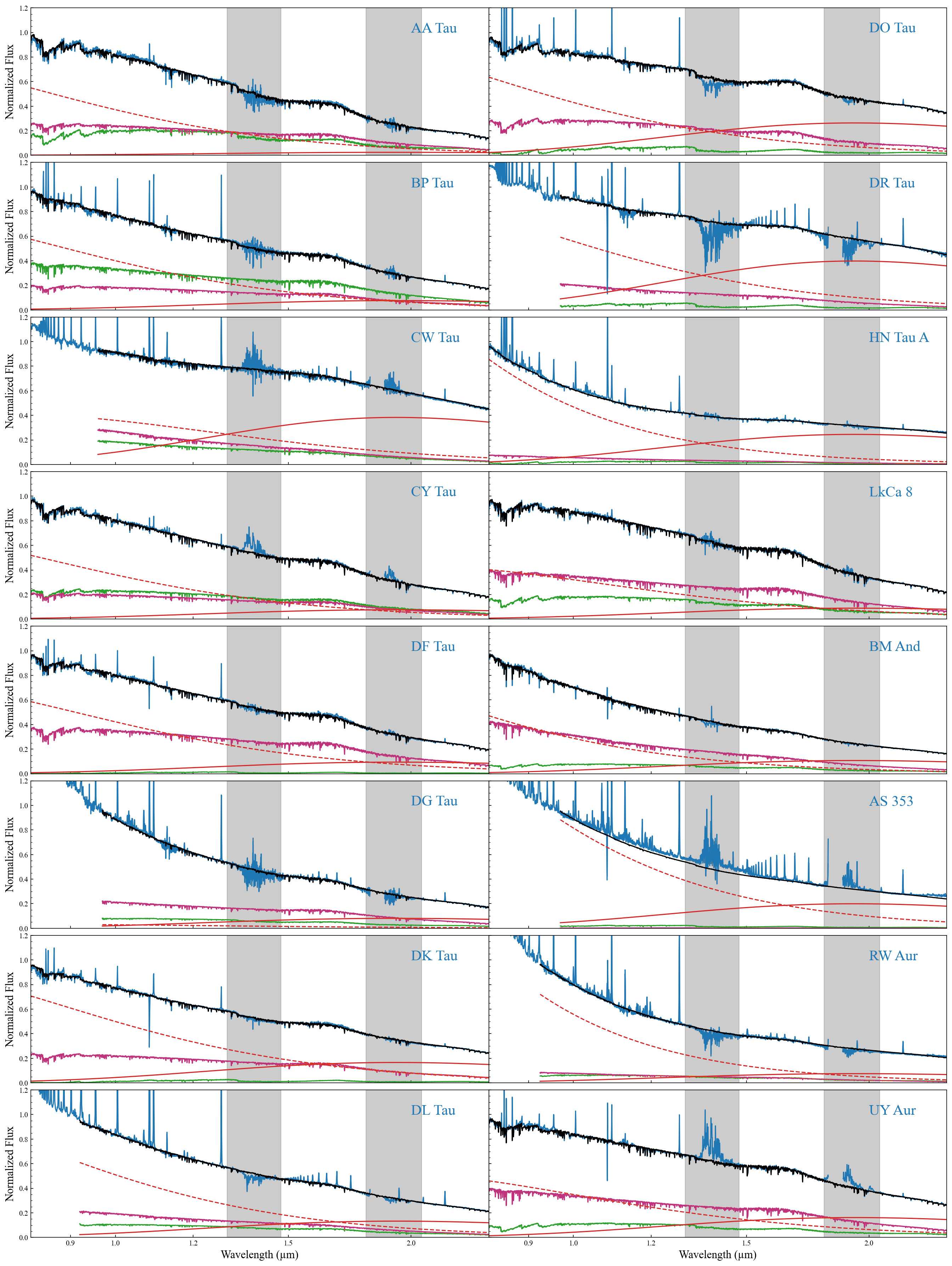}
\caption{{Multi-component spotted fits for all stars in our sample.} Colors are the same as in Figure~\ref{fig:bp_tau_fit}, and plotted are individual $T_{phot}$, $T_{spot}$, $T_{disk}$, and $T_{acc}$ components}
\label{fig:all_fits}
\end{figure*}

\begin{figure*}
\centering
\includegraphics[width=1\linewidth]{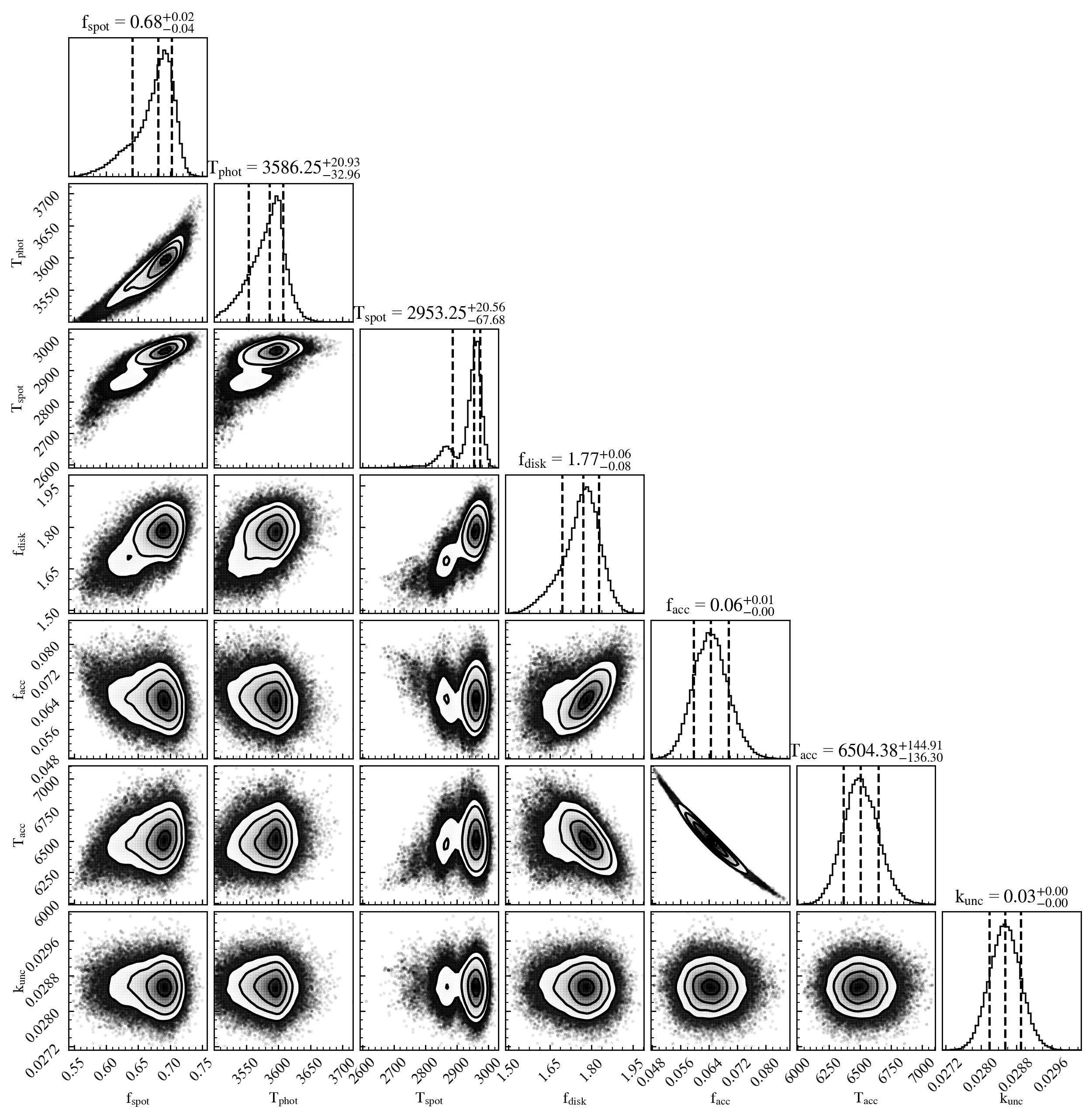}
\caption{{Corner plot for the spotted temperature accreting fit for AA~Tau. Overplotted on the histograms are kernel density estimation curves (solid black lines) as well as median and one-sigma percentiles (dashed black lines).}}\
\label{fig:aa_tau_fit_posterior}
\end{figure*}

In \citet{PerezPaolino2024}, we developed spectral models of spotted but non-accreting WTTS by combining two synthetic spectra from BTSettl-CIFIST \citep{allard2014} theoretical atmospheres of different temperatures over the $0.75-2.40\ \micron$ spectral range. We combined a hotter template representing the non-spotted photosphere, $F_{\lambda}(T_{phot})$, with a cooler one representing the spots, $F_{\lambda}(T_{spot})$, weighted by a spot filling factor, $f_{spot}$, that corresponds to the fraction of the visible stellar hemisphere covered in spots at the time of observation. In order to account for excess emission from warm dust at the inner edge of the disk, we added continuum emission from a blackbody, $B_{\lambda}(T_{disk})$, with $T_{disk}$ set to 1500~K {\citep{Pollack1994}} and a disk filling factor, $f_{disk}$. {In this study of accreting CTTS, we also include continuum emission from a second blackbody $B_{\lambda}(T_{acc})$ with temperatures ranging between 4000-12500~K \citep{Kwan2011}} and filling factor $f_{acc}$ to account for excess emission from the accretion activity. Our models of young spotted stars actively accreting material from a circumstellar disk therefore are of the following form:


\begin{equation}
\begin{split}
    F_{\lambda,\ model} = F_{\lambda}(T_{phot})(1-f_{spot}) + F_{\lambda} (T_{spot})f_{spot}  \\ +\ \pi (B_{\lambda}(T_{disk})f_{disk} + B_{\lambda}(T_{acc})f_{acc})
\end{split}
\end{equation}

{The temperatures and filling factors can be highly variable across time, as is expected from the observed spectroscopic \citep{myself} and photometric \citep{Rebull2020, Rebull2022UCL, Cody2022} variability for WTTS of different ages, and for accreting CTTS \citep{Bary2014} due to modulated accretion { activity}. Therefore, we emphasize the fact that these parameters are determined from fits to single observations and represent a snapshot of surface configuration of these stars. Changes in the total spot coverage of these stars are likely given typical spot lifetimes \citep{Basri2022} on the order of a few hundred days as seen in K2 photometry and observed multi-year lightcurve morphology changes in WTTS \citep{Gully2017}. We note that the spot-lifetimes in \citet{Basri2022} sample are likely much shorter than those of WTTS and CTTS \citep[e.g.,][]{Stelzer2003}.}

The fit parameters were constrained using an \textit{emcee} \citep{Foreman-Mackey2013} based Markov-Chain Monte-Carlo (MCMC) algorithm allowing for maximum likelihood estimation (MLE) and posterior analysis. In this way, we simultaneously fit for the non-spotted photospheric, spot, and accretion temperatures, as well as for the spot, accretion, and disk filling factors. The MCMC fits used 50 walkers with 5500 steps each and a 500 step burn-in. Regions of poor atmospheric transmission between $J$ and $H$-band (1.30--1.45 $\micron$) and between $H$ and $K$-band (1.80--2.05 $\micron$) were masked during fitting, along with the first and last 200 $\AA$ of the SpeX spectra to avoid areas of poor SNR at the edges of the detector. Additionally, regions containing emission lines were masked out during the fitting as our synthetic spotted star models do not account for line emission associated with accretion activity. Spot filling factors were constrained to the range 0--1, while accretion and disk filling factors were allowed to vary as free parameters. Temperature constraints for the spots and photospheres were set by the temperature coverage of the spectral models (1200~K--7000~K).

An additional parameter was added to our models in order to account for flux uncertainties in the BTSettl-CIFIST model atmospheres. This is represented by $K_{unc}$, or the fractional uncertainty in the model fluxes that is added in quadrature to the data uncertainty such that

\begin{equation}
    \sigma_{final} = \sqrt{\sigma_{data}^2+K_{unc}^2F_{\lambda,\ model}^2}
\end{equation}

\noindent This parameter was then minimized alongside the others in the MCMC algorithm. 

{Since the accretion shock is not entirely in local thermodynamic equilibrium or optically thick, the use of blackbody emission to model the accretion shock is a simple approximation. The presence of variable non-LTE line emission fills in neighboring stellar absorption lines and often times turns them to emission \citep{Dodin2013}, while Paschen and Balmer continuum emission lead to strong discontinuities or jumps in the spectrum depending on the density of the gas and the temperature of the shock \citep{Valenti1993}. Such features cannot be recreated by a simple blackbody. However, we can use blackbodies as appropriate approximations for the accretion excess because our spectral coverage begins at $0.8\ \micron$, the short-wavelength cutoff of SpeX. At these wavelengths, the accretion emission is dominated by the heated photosphere, which to a first-order approximation emits like a typical blackbody (See e.g. \citet{Gullbring1998, Ingleby2013}).} A full non-LTE analysis of the accretion characteristics of these stars is well beyond the scope of this work \citep[see e.g.,][]{Robinson2019}.

{In Figure~\ref{fig:bp_tau_fit}, we show an example of a two-temperature photosphere and starspot spectral model combined with emission from accretion hot spots and the warm inner disk for BP Tau. Similar fits for all 16 stars in our sample are shown in Figure~\ref{fig:all_fits}.} Typical root-mean-square (RMS) residuals for all fits across the entire sample are on the order of 0.03 or 3\% of the observed fluxes. We note that prior to fitting, we dereddened all of the spectra using the $A_V$ values reported by \citet{Fischer2011} in order to provide a direct comparison of the extracted excesses (Section~\ref{sec:excess}).

{A close inspection of the fits shown in Figure~\ref{fig:bp_tau_fit} and Figure~\ref{fig:all_fits} reveal notable spectral features that originate primarily in the starspots. While both the photosphere and the starspots contain TiO absorption bands at the temperatures found here, these are much stronger in the spectra of the cooler starspots. Given the large filling factors of the spots, 
the presence of strong TiO absorption features in the spectra can be attributed primarily to their heavily spotted nature and not to a photospheric origin. Similarly, the hump in the H-band is much stronger in the spectra of the starspots than in the photospheric spectrum. {This region can therefore likely serve as a reliable spot-indicator in WTTS with optical K- spectral types that should not have such a pronounced hump}. The combination of the steeper continuum from a hotter photosphere with cooler TiO and continuum from the spots leads to an {\it apparent} near-infrared excess, which is one of the hallmark signs of starspots on young stars.}

{A typical posterior for our model fit to AA~Tau is depicted by a corner plot shown in Figure~\ref{fig:aa_tau_fit_posterior} along with overplotted median and one-sigma percentiles. Our posteriors for all fits show single-peaked distributions indicative of well constrained fit parameters. We find no evidence for degeneracies in our fit temperatures or filling factors, as was also the case in \citet{myself, PerezPaolino2024}. We note however that there exists a strong anticorrelation between the accretion filling factor and temperature. This is likely caused by the fact that we are observing the Rayleigh-Jeans tail of the blackbody function for this hotter component, and its slope is not very well constrained at these wavelengths. Both a cooler accretion blackbody with a larger filling factor, and a hotter one with a smaller filling factor, can provide the same excess.}  

\subsection{Moderately Veiled Stars: Typical Spotted Photospheres for CTTS}
Across our sample we see spot filling factors between 45\% and 87\% of the stellar surface, with photospheric and spot temperatures between $3304\leq T_{phot} \leq 5667~K$ and $2189 \leq T_{spot} \leq 3257~K$. The median spot-to-photosphere temperature ratio is $0.77\pm{0.12}$, which agrees with ratios measured by \citet{Herbst2020} for active giants, sub-giants and dwarfs with non-spotted photospheric temperatures between 3000 and 4000~K, as well as with our previous results on WTTS \citep{PerezPaolino2024}. The resulting fit parameters for all stars in our sample are shown in Table~\ref{tab} where we have also computed spot-corrected effective temperatures following \citet{Gully2017} as

\begin{equation}\label{eq:efftemp}
    T_{eff} = \left[T_{phot}^4(1-f_{spot})+T^4_{spot}f_{spot}\right]^{1/4}
\end{equation}

{Comparing spot-corrected effective temperatures to the optical temperatures provided in \citet{Fischer2011}, we see temperature shifts on the order of 700~K. For this comparison, we have translated the spectral types reported in \citet{Fischer2011} to effective temperatures using the temperature conversions of \citet{Pecaut2013}. These shifts are significant, corresponding to an entire spectral type for the most heavily spotted sources. We note that the photospheric temperature returned by our fitting routine is highly sensitive to the assumed visual extinction to the source, as was shown in \citet{PerezPaolino2024}. We have adopted the same extinctions as \citet{Fischer2011} in order to provide a direct comparison of the measured excesses.
}

\begin{figure*}
\centering
\includegraphics[width=1\linewidth]{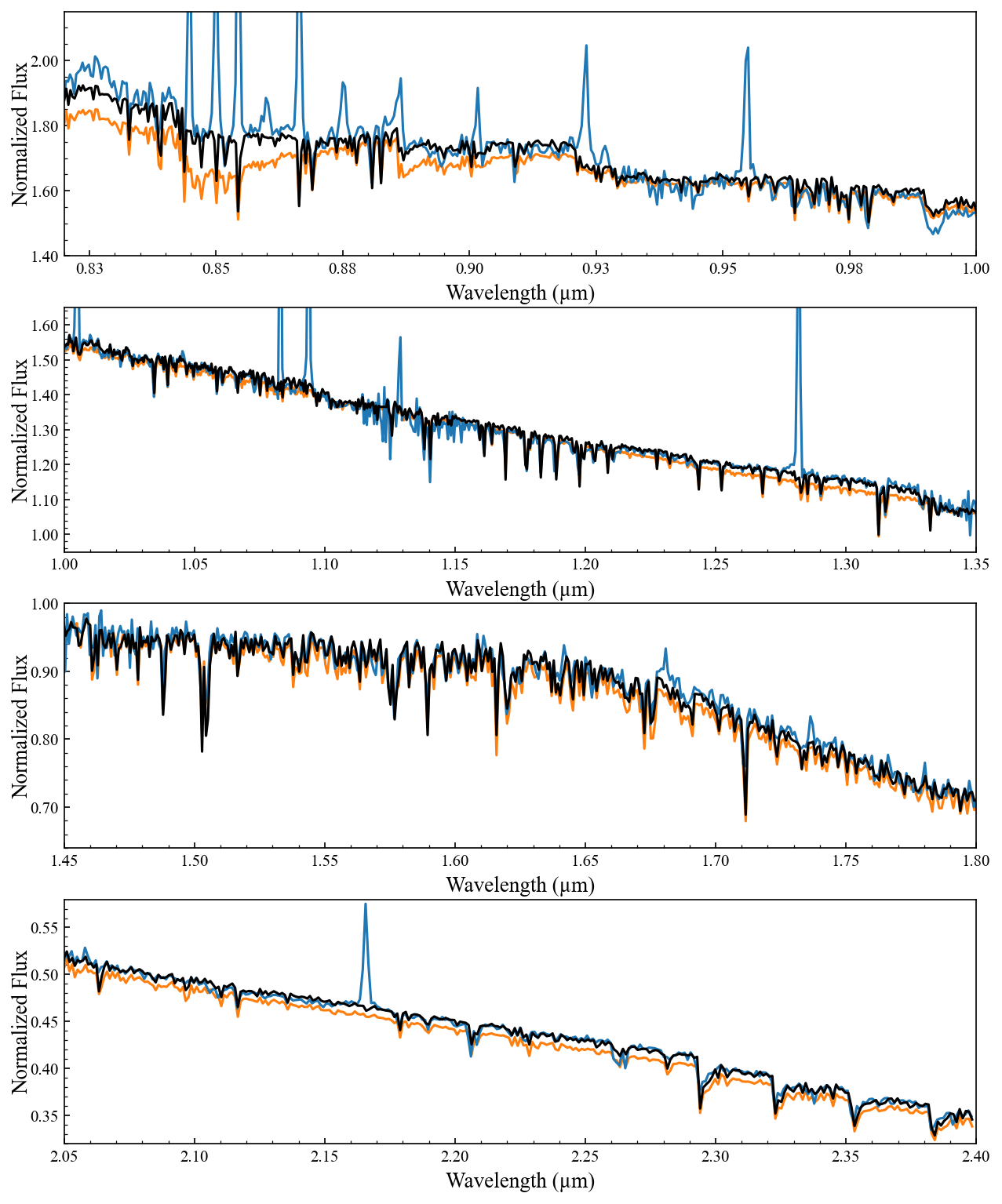}
\caption{{Comparison between spotted and unspotted model fits for BP~Tau. The dereddened spectrum is shown in blue, while the best fit spotted model is in black and unspotted model is in orange. The panels from top to bottom show IY-, J-, H-, and K-band spectra. The unspotted or single-temperature model less accurately predicts the continuum in certain wavelength ranges, as well as line and molecular band strengths.}}
\label{fig:closeup_tau_fit}
\end{figure*}

{
Assuming that starspots on CTTS behave like those on WTTS, we expect to see at most a systematic shift of 300~K in our photospheric temperatures.
But the inferred temperature shifts are larger, around 700~K,
y
when adopting spotted models of accreting stars. We demonstrate in Figure~\ref{fig:closeup_tau_fit} the improvement over models of accreting stars that do not incorporate starspots. Shown is a comparison between a spotted accreting fit (same as that in Figure~\ref{fig:bp_tau_fit} but zoomed in to highlight the spectral features) and a single temperature fit for BP~Tau. For the single-temperature fit, we eliminated the spot component, fitting BP~Tau as a stellar photosphere with superimposed blackbody accretion and disk excesses. Our best-fit single temperature model has a photospheric temperature of $T_{phot}=3525~K$, a disk filling factor of $f_{disk}=12.5$ times the stellar surface area, and an accretion temperature of $T_{acc}=5057~K$ with a filling factor of $f_{acc}=0.278$.
These can be compared to the spotted accreting fit parameters of
$T_{phot}=3939~K$ for the star, $f_{disk}=9.24$ for the cool fixed-temperature disk component, and $T_{acc}=6736~K$ and $f_{acc}=0.068$ for the hot accretion component, where the additional starspot had $T_{spot}=2699~K$ and $f_{spot}=0.51$. 
}

{
BP~Tau has an optical spectral type of K7, necessitating the presence of emission at $\approx4000~K$, which is provided by the accreting spotted model fit but not the single-temperature model fit.
Given the sharp drop in the contribution of the starspots to the stellar spectrum compared to that of the photosphere for wavelengths shorter than $\approx0.75\ \micron$ \citep{PerezPaolino2024}, we expect optical spectral types to most closely match the photospheric temperature of a spotted star. BP~Tau cannot have a single-temperature photosphere at 3525~K while displaying a K7 spectral type at optical wavelengths. Additionally, this temperature would be too low for an unspotted star with a dynamical mass of $1.1\ M_{\odot}$ \citep{Simon2019} for all commonly accepted ages for Taurus-Auriga \citep[see models of e.g.,][]{Baraffe2015}. If instead we assume that the star is spotted with spots covering 51\% of the stellar surface, then we can reconcile the optical spectral types with the photospheric temperature and spot-corrected effective temperature of 3471~K (Table~\ref{tab}).}

{While the single-temperature fit matches the continuum level of the observed spectrum at H-band (Figure~\ref{fig:closeup_tau_fit} panel 3), it fails to reproduce the spectrum for wavelengths shortward of $\approx0.95\ \micron$ and predicts TiO absorption bands at $0.85\ \micron$ and $0.89\ \micron$ that are far too strong. The spotted model on the other hand better matches the strength of these features while reproducing the continuum level. A similar story is seen at K-band, where the spotted model accurately replicates the continuum level and strength of the CO bands at $\approx2.3\ \micron$.}

\subsection{Highly Veiled Stars: Constraints on Photosphere Parameters}
AS~353~A, DR~Tau, CW~Tau, DG~Tau~A, DL~Tau, DR~Tau, and RW~Aur~A are CTTS with high mass accretion rates \citep[e.g.,][]{Gangi}. Therefore, the fitting range had to be adjusted due to either the presence of a Paschen jump that blackbody emission alone could not reproduce, or contamination from Paschen series emission lines obscuring the stellar continuum. As a result, the spectral region short of 0.95 \micron\ was removed from the fitted regions, which previous studies have found to be very sensitive to starspots and is important for constraining the temperature of the spots during spectral fitting \citep{Gully2017, myself, PerezPaolino2024}. 

\begin{figure}
\centering
\includegraphics[width=1\linewidth]{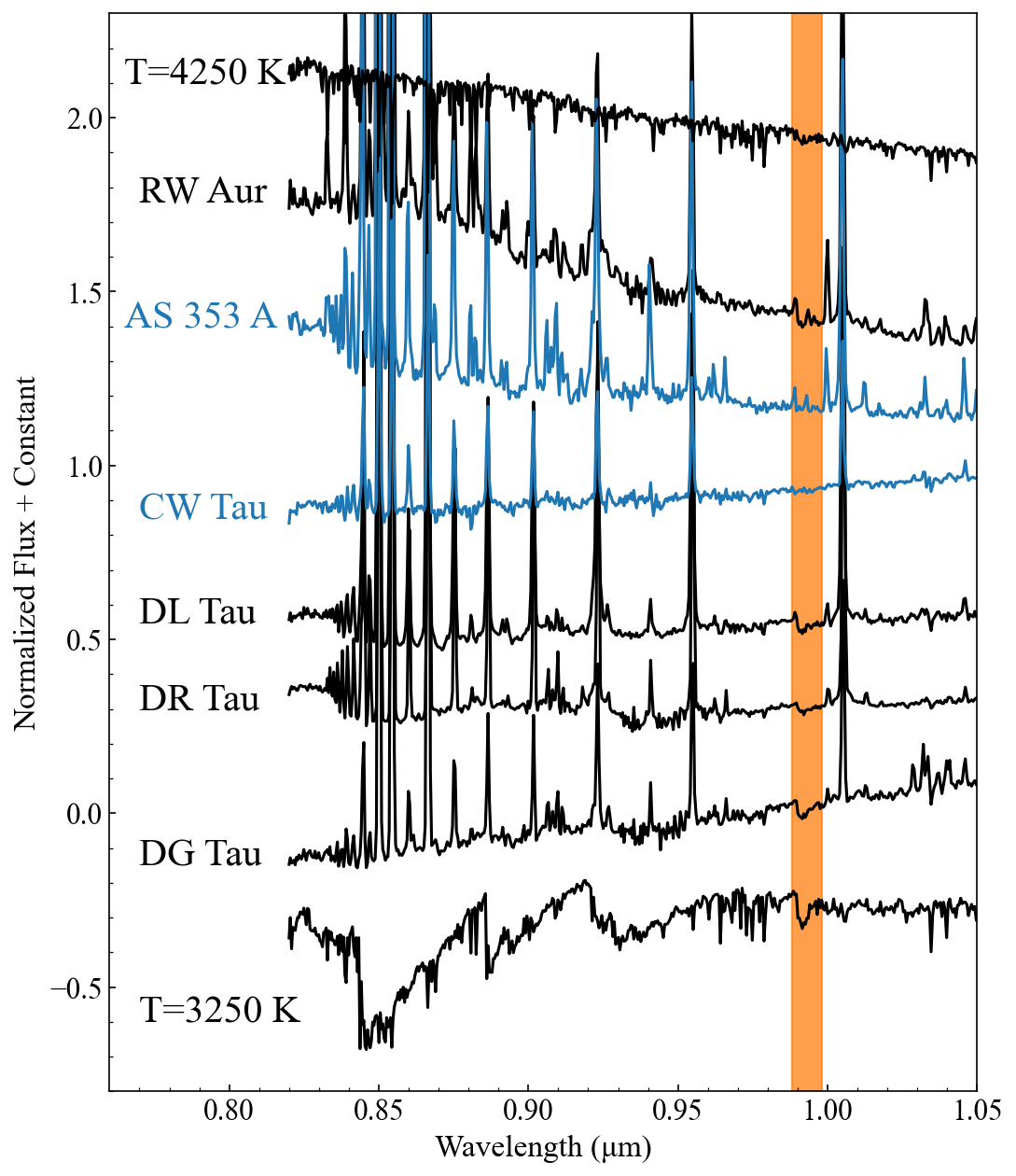}
\caption{{Presence of FeH in highly veiled stars.} Shown is a comparison of the spectral region between 0.80--1.05 $\micron$ for the highly veiled stars in our sample. Stars plotted in blue do not show evidence for the presence of FeH at 0.99 $\micron$ (region shaded in orange) in their spectra, while those plotted in black do show FeH absorption. Also plotted are a 3250~K and a 4000~K theoretical atmosphere for comparison.} 
\label{fig:feh}
\end{figure}

{DG~Tau~A and AS~353~A were fit using the extinctions found by \citet{Fischer2011}. However, we find that the accretion temperature across all our fits for both of these objects returns an unrealistically high 12500~K, the largest value allowed during the fitting. We therefore tried to refit these targets using lower values from the literature of $A_V=2.1$ and $A_{V}=1.6$ from \citet{Hartigan1995} and \citet{herczeg2014}, respectively, as we were concerned that those from \citet{Fischer2011} may have been overestimated. Despite this change, the accretion temperature again returned 12500~K. We believe that this may have been caused by an overestimated extinction in \citet{Fischer2011} leading to a steep spectral slope that could only be reproduced with unrealistically high accretion temperatures. Since we saw no change in the accretion temperatures for different values of $A_V$, in this paper we present fit parameters using the $A_V$ values reported in \citet{Fischer2011} in order to allow for a direct comparison with the excesses presented therein.}

In order to provide stronger temperature constraints for our heavy accretors, we looked for the presence of FeH absorption at $0.99~\micron$. Due to its low dissociation energy of 2.9~eV \citep{Wang1996}, it will only exist within the coldest regions of the stellar surface. Previous studies of K- and M-type dwarfs have found FeH to be highly temperature and gravity sensitive \citep{rayner2009}, being strongest for late M-type dwarfs with decreasing strength for higher temperatures until disappearing entirely by K5 {\citep[4400~K,][]{Pecaut2013}}. Therefore, if a CTTS spectrum has FeH absorption, we can either place an upper bound on the photospheric temperature or infer the presence of spots. Additionally, given location of the 0.99$\micron$ FeH feature in the spectrum, it is not contaminated by any strong emission lines, making it a more reliable spectral type indicator for our sample. 

We separated our sample of heavy accretors into those that showed FeH absorption (RW~Aur~A, DL~Tau, DR~Tau, and DG~Tau~A) and those that did not (AS~353~A and CW~Tau) as shown in Figure~\ref{fig:feh}, where we have also included two synthetic atmospheres of 3250~K and 4250~K for comparison. For stars with detectable FeH absorption, we tend to see high spot filling factors for spots with temperatures under 3400~K (Table~\ref{tab}) that have photospheres too hot to account for the presence of FeH. Therefore, while the actual spot filling factors or temperatures may be unreliable due to the difficulties associated with fitting heavily accreting stars in this manner, or an erroneous $A_V$ value, we can confidently point to the presence of cooler material on the stellar surfaces likely associated with spots.

For the two stars without FeH, AS~353~A and CW~Tau, our results are more complicated. Fits for CW~Tau point to an unspotted star with a photospheric temperature on the order of $\approx4700~K$, which is consistent with the lack of FeH in its spectrum and previous spectral type estimates \citep[K3, ][]{Kenyon1995}. In the case of AS~353~A, fits were particularly poor due to the extreme contamination from accretion emission lines and continuum veiling that lead to the lack of any discernible spectral features. In this case, the temperature of the photosphere and that of the spots converged to the same value, indicating an unspotted star and leading to an unconstrained spot filling factor. We cannot rule out the presence of starspots based solely on this result, and it is worth noting that the photospheric and spot temperatures of $\approx 2400~K$ are far too low to agree with previous spectral type estimates for this source \citep[K5, ][]{Hartigan1995}. {Additionally, the lack of FeH in its spectrum rules out such a low photospheric and spot temperature because if the star were that cool, we should see strong FeH absorption. We credit the failure of our fitting routine in constraining the surface temperatures of AS~353~A to the extreme line emission and continuum veiling in this source, { as well as to the possibility of an overestimated extinction value.} We include AS~353~A in Figure~\ref{fig:all_fits} and Figure~\ref{fig:all_excess_comparison} to show the limitations of our model, and do not believe these to be realistic estimates of the spectral components or excesses.}

\section{Deriving the Excess Emission Spectrum}\label{sec:excess}
Following a successful spectral decomposition, we now characterize the spectra of the continuum excess superimposed on the intrinsic stellar spectrum using our spotted models and compare our results to those of \citet{Fischer2011}. Using the best-fit spot and photosphere with their corresponding filling factors, we subtract the stellar contribution from every de-reddened SpeX spectrum to estimate the excess emission $E_{\lambda}$ coming from the accretion hot spots and the disk, where

\begin{equation}
    E_{\lambda} = F_{\lambda,\ obs} - (F_{\lambda}(T_{phot})(1-f_{spot})+F_{\lambda}(T_{spot})f_{spot}) 
\end{equation}

\begin{figure*}
\centering
\includegraphics[width=1\linewidth]{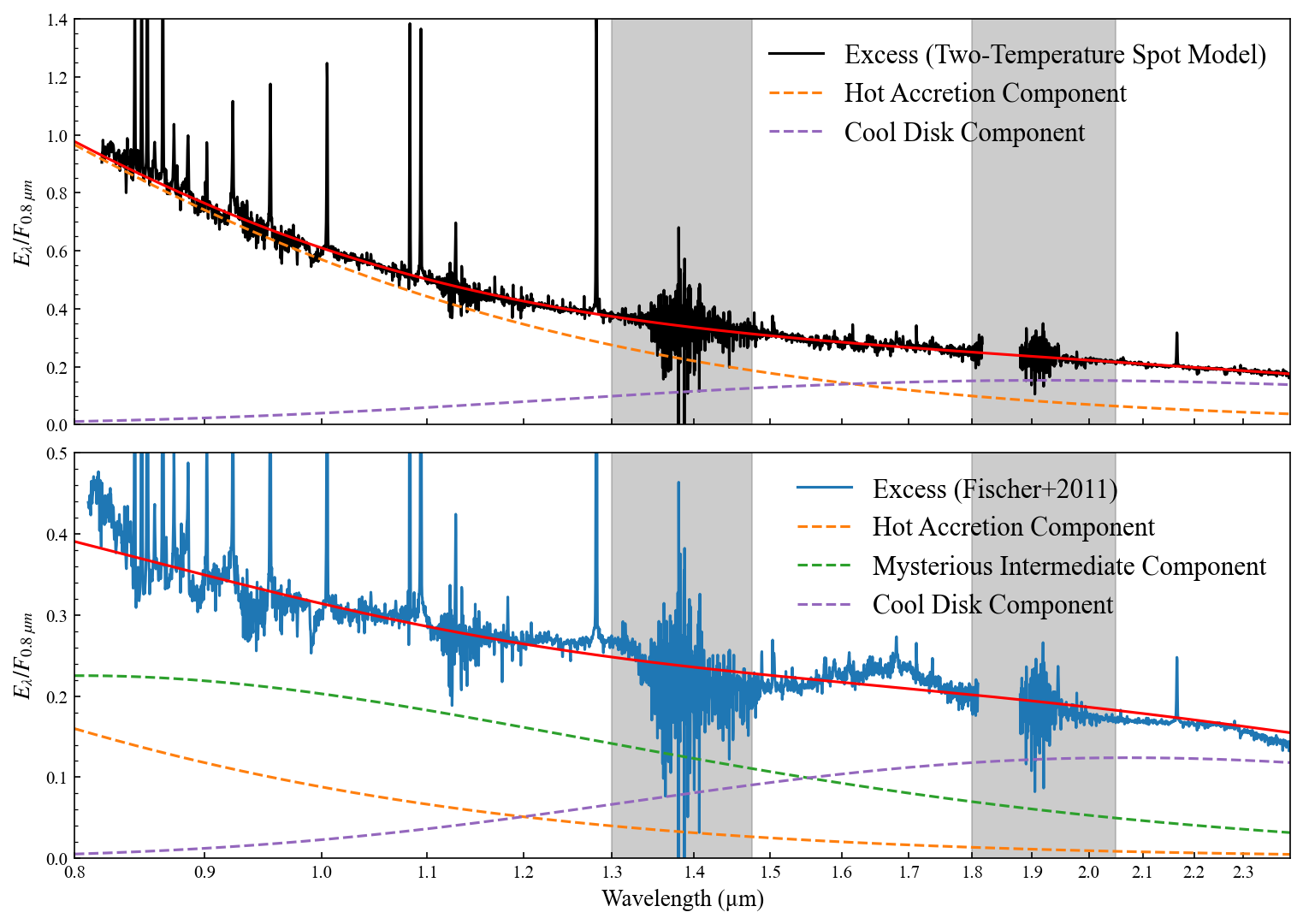}
\caption{{Excess spectra for BP~Tau}. (Top) Plotted is the excess emission across the SpeX spectral coverage (black solid line) calculated from the two-temperature spotted fits to the central star. Alongside this are the best-fit disk (purple dashed line) and accretion (orange dashed line) components that make up the excess. (Bottom) Plotted is the same comparison, however, we now consider the excess as calculated in \citet{Fischer2011} assuming a single-temperature photopshere (blue solid line) and include the third mysterious intermediate temperature component (dashed green line) alongside the disk (dashed purple line) and accretion (dashed orange line) components necessary to reproduce the excess. The top panel with the spotted photosphere provides a more satisfactory fit to the spectrum. Regions of poor atmospheric transmission that were masked during the fitting process are shaded in grey.}
\label{fig:excess_comparison}
\end{figure*}

\begin{figure*}
\centering
\includegraphics[width=1\linewidth]{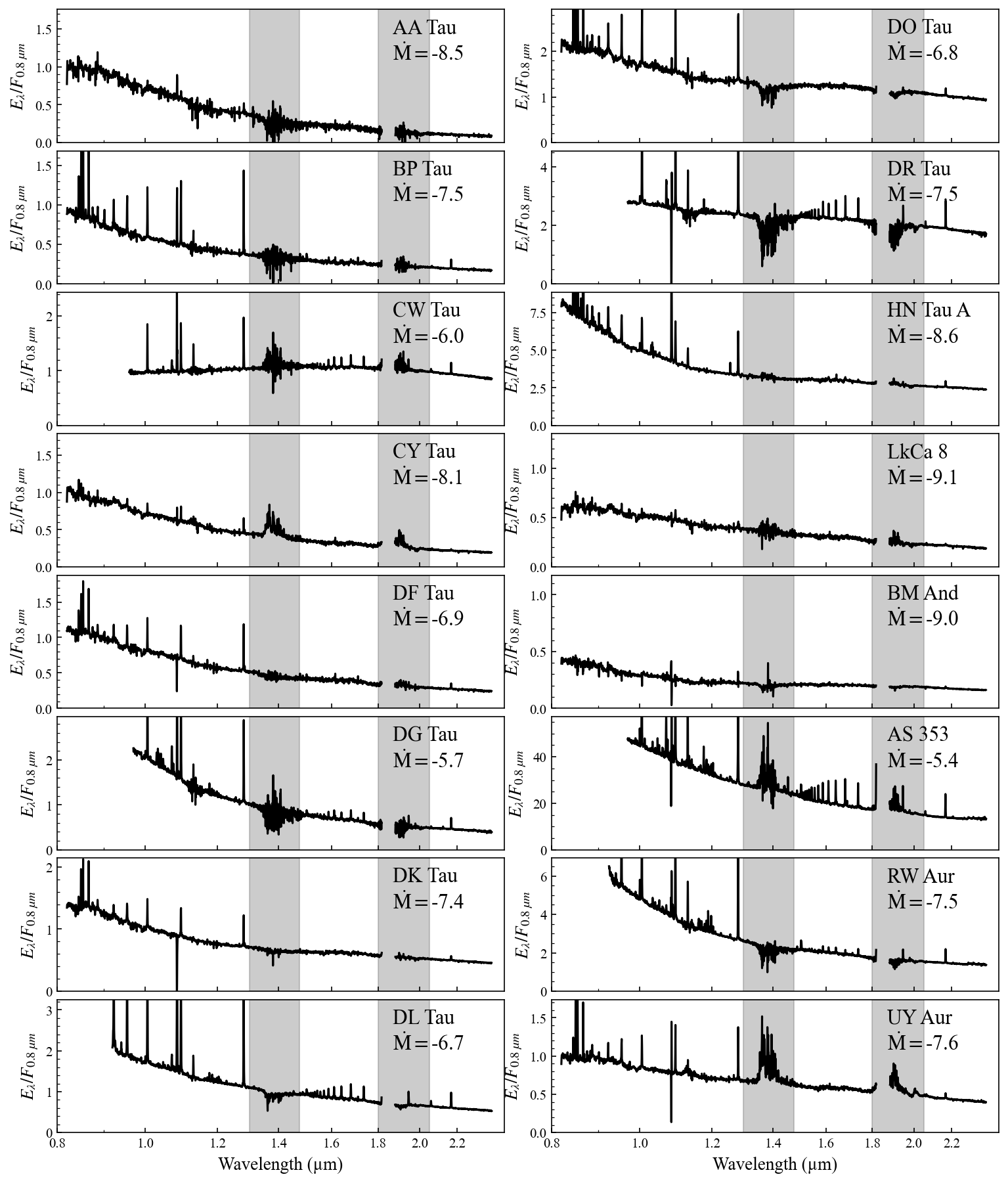}
\caption{{Excess emission spectrum for all stars in our sample.} Emission spectra were calculated from the best-fit parameters of the two-temperature accreting fits, while the listed mass accretion rates are from \citet{Fischer2011}.}
\label{fig:all_excess_comparison}
\end{figure*}

\noindent In the top panel of Figure~\ref{fig:excess_comparison}, we show an excess spectrum for BP~Tau, normalized to the intrinsic stellar flux at 0.8 \micron, along with the individual disk and accretion components that contribute to the excess. clearly, the excess shows a monotonic increase towards shorter wavelengths, as is expected of an accreting CTTS, with the disk dominating the excess for wavelengths longer than $\lambda\geq1.5\ \micron$. The excess spectra for all stars in our sample are shown in Figure~\ref{fig:all_excess_comparison}. 

In \citet{Fischer2011} the excess spectrum of BP~Tau was calculated by first using V819~Tau as a spectral standard to measure veiling across ten regions in the NIR SpeX spectra. Then, these veiling measurements were used to scale main sequence spectral templates and subtract the intrinsic stellar spectrum from the observed CTTS spectra. As main-sequence templates, \citet{Fischer2011} used dwarfs from the SpeX IRTF library \citep{rayner2009}. In the bottom panel of Figure~\ref{fig:excess_comparison}, we show the resulting excess spectrum for BP~Tau obtained using the method of \citet{Fischer2011} and  their veiling measurements. Using a blackbody at 8000~K to represent the accretion excess and a blackbody at 1400~K to represent the disk, \citet{Fischer2011} attempted to fit their excess spectra to determine the relative sizes of the emitting regions. However, their results suggested the need for an additional intermediate temperature component with a temperature between 2200--5000~K. Overplotted in Figure~\ref{fig:excess_comparison} are the three best fit components that make up the excess for BP~Tau, showcasing the need for the intermediate temperature component in order to reproduce the shape of the excess, particularly around 1.0--1.2 \micron. The size of this component relative to the stellar photosphere required to reproduce the excess varies from star to star, spanning a range between 14\% and eleven times the stellar surface area across all sixteen CTTS. 

While the introduction of this third component provided a better fit to the overall shape of the excess, their (blackbody) fits were still unable to reproduce the observed excesses. A close inspection of the excess spectrum for BP~Tau shown in {the bottom panel of} Figure~\ref{fig:excess_comparison} reveals notable features that are still not fit well, the largest of which are several strong molecular absorption bands that most closely resemble those found in the spectra of M-type dwarf stars and that could not have originated in the accretion shock or disk. The most significant of these are the TiO bands located between 0.80 and 1.00 \micron, the sodium doublet at 0.82 \micron, the FeH feature at 0.99 \micron, the water absorption band at $\approx$1.30 \micron, the hump in the H-band, and the CO bands at $\approx2.3\ \micron$. This trend of molecular features being present in the excess is seen across all stars in the \citet{Fischer2011} sample.

In the case of BP~Tau, \citet{Fischer2011} used a K7V main sequence dwarf template to subtract the stellar component from their observed spectrum. Assuming that our best-fit spotted star model provides a better fit to the underlying stellar spectrum of BP Tau, then subtracting only a K7 dwarf template would subtract the photospheric contribution only, leaving behind residuals approximating the spectrum of an M-type star with strong molecular features. For IYJ wavelengths where the intermediate temperature component dominates, emission from the starspots can surpass the photospheric contribution and become the dominant source, explaining the relative strength of the IYJ excess in this region and the presence of molecular features in the excess. Given that the emission from starspots with temperatures similar to those of an M-type star would be strongest at wavelengths between 0.80--1.50 \micron\ and also provide the necessary spectral features, the presence of large cool starspots may replace the need for the mysterious intermediate temperature component suggested by \citet{Fischer2011}. 


Inspecting the excess computed using our models of spotted accreting stars {(see top panel of Figure~\ref{fig:excess_comparison})}, we find that while some of the spectral features are not entirely eliminated from the excess, their scale is greatly reduced indicating a more accurate subtraction of the intrinsic stellar spectrum. Most importantly, the shape of the excess now resembles a smooth continuum instead of a stellar spectrum, and the shape can be reproduced entirely using only two additional (blackbody) components; one for the accretion and another for the disk. This picture is repeated for all stars in our sample regardless of spectral type or accretion rate, as can be seen in Figure~\ref{fig:all_excess_comparison} where we have normalized the excess spectrum by the intrinsic stellar flux at 0.8~\micron. Shown are also mass accretion rates for every star from \citet{Fischer2011}, suggesting that there exists a correlation between mass accretion rates and the scale of the excess relative to the stellar flux at 0.8~\micron\ across all stars in our sample. 

Given these results, we conclude that the introduction of starspots into the stellar component eliminates the need for an intermediate temperature component when fitting the excess. We therefore attribute its proposed existence in the literature to the unaccounted spotted nature of these stars. This result highlights the importance of the starspot spectrum even in CTTS, for which accretion is typically assumed to be the dominant excess source.

\begin{deluxetable*}{lcccccccccc}
\tablecaption{Best-Fit Two-Temperature Spotted Parameters.}\label{tab}
\tablehead{
\colhead{Star} & \colhead{SpT$^{(1)}$} & \colhead{$A_{V}^{(1)}$} & \colhead{$f_{spot}$} & \colhead{$T_{phot}$ [K]} & \colhead{$T_{spot}$} & \colhead{$f_{disk}$} & \colhead{$f_{acc}$} & \colhead{$T_{acc}$} & \colhead{$K_{unc}$} & \colhead{$T_{eff}$}\\
\colhead{} & \colhead{} & \colhead{(mags)} & \colhead{} & \colhead{(K)} & \colhead{(K)} & \colhead{} & \colhead{} & \colhead{(K)} & \colhead{(K)} & \colhead{(K)}} 
\startdata
AA Tau & K7 & 1.34 & $0.68\pm{0.08}$ & $3599\pm{8}$ & $2990\pm{46}$ & $1.92\pm{0.08}$ & $0.060\pm{0.008}$ & $6728\pm{183}$ & $0.026$ & $3223\pm{58}$\\
AS 353 A$^{(2)}$ & K7 & \nodata & $0.71\pm{0.36}$ & $2416\pm{125}$ & $2406\pm{110}$ & $132.28\pm{21.4}$ & $0.976\pm{0.200}$ & $6957\pm{83}$ & $0.028$ & $2409\pm{86}$\\
BM And A & G8 & 1.60 & $0.65\pm{0.04}$ & $5667\pm{151}$ & $3257\pm{287}$ & $30.40\pm{1.64}$ & $0.076\pm{0.040}$ & $7399\pm{466}$ & $0.012$ & $4555\pm{155}$\\
BP Tau & K7 & 1.75 & $0.51\pm{0.02}$ & $3939\pm{31}$ & $2699\pm{48}$ & $9.24\pm{0.32}$ & $0.068\pm{0.016}$ & $6736\pm{324}$ & $0.014$ & $3471\pm{34}$\\
CW Tau$^{(2)}$ & K3 & 2.10 & $0.51\pm{0.04}$ & $4969\pm{30}$ & $4496\pm{22}$ & $153.88\pm{1.24}$ & $1.876\pm{0.040}$ & $3663\pm{20}$ & $0.009$ & $4745\pm{73}$\\
CY Tau & M1 & 1.19 & $0.44\pm{0.05}$ & $3594\pm{10}$ & $3208\pm{103}$ & $9.60\pm{0.24}$ & $0.104\pm{0.012}$ & $6096\pm{53}$ & $0.015$ & $3441\pm{42}$\\
DF Tau & M2 & 1.77 & $0.44\pm{0.05}$ & $3600\pm{10}$ & $3091\pm{64}$ & $11.92\pm{0.52}$ & $0.108\pm{0.020}$ & $6136\pm{179}$ & $0.023$ & $3405\pm{33}$\\
DG Tau A & K7 & 5.43 & $0.41\pm{0.13}$ & $3759\pm{88}$ & $3113\pm{109}$ & $23.52\pm{3.24}$ & $0.076\pm{0.056}$ & $12442\pm{1357}$ & $0.013$ & $3538\pm{104}$\\
DK Tau A & K7 & 1.83 & $0.75\pm{0.06}$ & $4139\pm{123}$ & $3069\pm{52}$ & $25.72\pm{1.84}$ & $0.108\pm{0.052}$ & $6644\pm{356}$ & $0.024$ & $3440\pm{97}$\\
DL Tau & K7 & 3.00 & $0.55\pm{0.07}$ & $4290\pm{45}$ & $3284\pm{75}$ & $48.36\pm{6.64}$ & $0.572\pm{0.188}$ & $5646\pm{107}$ & $0.013$ & $3836\pm{79}$\\
DO Tau & M0 & 3.04 & $0.50\pm{0.02}$ & $3337\pm{20}$ & $2281\pm{38}$ & $25.68\pm{0.48}$ & $0.124\pm{0.004}$ & $5730\pm{103}$ & $0.016$ & $2947\pm{25}$\\
DR Tau & K7 & 1.54 & $0.77\pm{0.11}$ & $4498\pm{44}$ & $2189\pm{140}$ & $71.6\pm{11.80}$ & $0.452\pm{0.044}$ & $4946\pm{287}$ & $0.020$ & $3265\pm{312}$\\
HN Tau A & K5 & 3.05 & $0.84\pm{0.01}$ & $5143\pm{108}$ & $2820\pm{46}$ & $160.08\pm{9.48}$ & $0.204\pm{0.048}$ & $11182\pm{460}$ & $0.016$ & $3600\pm{64}$\\
LkCa 8 & M0 & 0.47 & $0.53\pm{0.09}$ & $3801\pm{89}$ & $2971\pm{25}$ & $9.08\pm{0.20}$ & $0.184\pm{0.024}$ & $4455\pm{302}$ & $0.015$ & $3434\pm{92}$\\
RW Aur A & K1 & \nodata & $0.90\pm{0.06}$ & $4329\pm{168}$ & $3214\pm{88}$ & $72.00\pm{12.08}$ & $0.152\pm{0.080}$ & $12466\pm{881}$ & $0.020$ & $3377\pm{122}$\\
UY Aur A & M0 & 1.54 & $0.47\pm{0.01}$ & $3951\pm{160}$ & $3065\pm{190}$ & $25.92\pm{3.72}$ & $0.396\pm{0.108}$ & $4357\pm{23}$ & $0.013$ & $3614\pm{124}$\\
\enddata
\vspace{0.1cm}
(1)~From \citet{Fischer2011}. (2)~Best-fit two-temperature spotted model returns an unspotted star.
\end{deluxetable*}

\section{Discussion}\label{sec:conclusion}
The excess spectra shown here {(Figure~\ref{fig:all_excess_comparison})} and calculated by accounting for starspots when characterizing the stellar contribution to the observed spectrum vary from star to star. However, in all cases we find evidence for a smooth continuum-like shape that increases towards shorter wavelengths and decreases for longer wavelengths, as is expected for an excess originating from an accretion shock and disk. The magnitude of the excess varies across our sample from half of the stellar flux to well over ten times the stellar flux at 0.8 \micron, where it dominates the intrinsic stellar emission. 

One notable consequence of these results is that veiling between 0.8-1.4 \micron\ is not flat \citep[e.g.,][]{McClure2013} but instead rises monotonically across the NIR towards the visual. {We note that this is not seen in all of our targets, as for the heaviest accretors, the fitting range had to be cut short of $\approx1 \micron$. However, we expect this trend to continue as the accretion shock excess dominates the stellar emission.} 

Likewise, the magnitude of the excess at IYJ wavelengths when computed while accounting for starspots, is higher for all stars than that found in \citep{Fischer2011}, resulting in higher values of veiling that could significantly affect mass accretion rates. {These results are in line with the findings of \citet{Kidder2021}, where starspots were identified as a possible source for the excess veiling at H- and K-band from analysis of high-resolution spectra of CTTS in Taurus-Auriga. Unexpectedly high veiling at these wavelengths had been previously noted by e.g. \citet{Folha1999} and \citet{Johns-Krull2001}.  Our findings suggest that this effect might be more substantial at H- and K- wavelengths due to the relative contribution of the spots with respect to the unspotted stellar photosphere being maximized in this region.} Given the fact that the stars in the \citet{Fischer2011} sample span four orders of magnitude in mass accretion rates, similar findings for a sample of stars with seemingly common characteristics across Taurus-Auriga imply that these results hold for stars regardless of mass accretion rate. Future work modeling accretion rates from simultaneous UV-NIR spectra while taking into account starspots could test this result. 

The use of WTTS templates to measure extinction towards targets in Taurus-Auriga has been extensively used in the past. For instance, \citet{Gullbring1998} used WTTS templates to simultaneously estimate veiling and $A_V$ while accounting for the accretion excess using a slab model following \citet{Valenti1993}. Similar methodologies were employed by \citet{Hartigan2003} for sub-arc-second binaries, \citet{Fischer2011} as discussed here, and by \citet{herczeg2014}. \citet{McClure2013} follow a similar methodology using slab models of separate energy fluxes to account for accretion funnels of different densities. Most recently, \citet{Robinson2019} followed this approach using more complex accretion models that include line emission and a full non-LTE physical modeling of the accreting gas. 

The common thread among all of these studies is the use of WTTS spectral templates to represent the stellar spectral contribution. In light of the results shown here and those of \citet{PerezPaolino2024}, we must call into question the reliability of this method. {This is because not only do most WTTS seem to posses significant spot coverage on their surfaces, but also because the starspot emission relative to the photosphere is maximized in the NIR}. Given that starspots can dominate the photospheric contribution while having an entirely different SED, they will bias extinction estimates even if the accretion contribution is appropriately subtracted.

\section{Conclusion}
We model the NIR SpeX spectra of 16 CTTS using composite spectral models that incorporate emission from the stellar photosphere and starspots, as well as disk and accretion continuum emission. Using a Markov-Chain Monte-Carlo algorithm, we constrain the sizes and temperatures of all the components, allowing us to reproduce the 0.80-2.40 \micron\ spectra of all our targets. We then derive the continuum excess emission and demonstrate that it can be reproduced by the simple combination of the well-established accretion hot spots and warm dust in the inner disk. {These results suggest that the mysterious intermediate-temperature component of the IYJ excess found in \citet{Fischer2011} may have been the misidentified signal of starspot emission, and not warm annuli around regions of the shock-heated photosphere, gas inside the dust sublimation radius, or gas in the accretion flow as was suggested therein.} 

As our understanding of the effects of starspots on young stars increases, the idea that the NIR is an ideal place for spectral type determinations of young stars must be put into question. While it is true that at NIR wavelengths the veiling caused by the accretion shocks and disk is minimized, it is also here where the contribution of starspots is maximized relative to that of the stellar photosphere. Therefore, any single spectral region may be insufficient to completely disentangle the emission from young stars without the need for complimentary spectroscopy at different wavelengths.

Given the seemingly ubiquitous presence of starspots on young stars, future studies of the youngest star forming regions will require a simple and reliable method to estimate spot contributions. As was found here, medium resolution spectroscopy is more than adequate for this task, allowing for future studies with shorter observational investments. 

We thank the anonymous referee for comments that helped us improve the presentation of our work. This paper is dedicated to the memory of our collaborator and friend Will Fischer.

\software{astropy \citep{astropy2013}, SpexTool \citep{cushing2004},matplotlib \citep{Hunter2007}, Scipy 
\citep{jones2014}, Numpy \citep{VanderWalt2011}, emcee \citep{Foreman-Mackey2013}}

\pagebreak
\bibliography{spots_working.bib}{}
\bibliographystyle{aasjournal}
\end{document}